\documentstyle[aps,prl]{revtex}
\begin{document}
\draft

\title{Effect of spatial variations of superconducting gap on 
suppression of the transition temperature by impurities}

\author{M. E. Zhitomirsky \cite{Landau}\  
and M. B. Walker}

\address{Department of Physics, University of Toronto, Toronto, 
Canada M5S 1A7}
\date{February 4, 1998}
\maketitle

\widetext 
\begin{abstract}
We calculate correction to the critical temperature of a dirty 
superconductor, which results from the local variations of the gap 
function near impurity sites. This correction is of the order of 
$T_c/E_F$ and becomes important for short-coherence length 
superconductors. It generally reduces a pair-breaking effect. In 
$s$-wave superconductors small amounts of nonmagnetic impurities can 
increase the transition temperature. 
\end{abstract}

The effect of disorder is of considerable significance for the physics 
of superconductivity. The well known picture for conventional $s$-wave 
superconductors includes a strong pair-breaking role of magnetic 
impurities \cite{AGm} versus a weak effect from time-reversal symmetric 
perturbations such as nonmagnetic impurities and other lattice defects 
\cite{AGnm}, in agreement with the Anderson's theorem \cite{Anderson}. 
Generalization to the case of anisotropic Cooper pairing shows that 
even nonmagnetic scatters are pair breakers for a momentum dependent 
gap. Such a behavior was observed for heavy-fermion and copper-oxide 
superconductors and the $d$-wave superconducting gap of the latter is 
now well established \cite{dwave}. The original Abrikosov-Gor'kov (AG) 
theory was extended to calculate changes of the transition temperature 
$T_c$, superfluid density, and residual density of states  for 
arbitrary strength of potential scatters in $d$-wave superconductors 
\cite{AGext}. 

On the experimental side, while some measurements on Zn-substituted 
Y-123 samples show perfect agreement with the AG theory \cite{exp1}, 
others observe a substantially slower-than-predicted reduction of 
$T_c$ with increasing impurity concentration in Y-123 and La-214 
systems \cite{exp2} or with growing ion damage on irradiated samples 
\cite{exp3}. Various theoretical explanations were proposed for this 
discrepancy including strong-coupling effects \cite{Radtke} or an 
effect from the van Hove singularity \cite{Fehr}. An alternative 
hypothesis was suggested by Franz and co-workers \cite{Franz}, who 
presented numerical evidence for the effect of spatial variations of 
the gap neglected in the AG theory on thermodynamic properties for a 
model two-dimensional $d$-wave superconductor. First discussions of 
such effects in conventional superconductors with magnetic impurities 
date to late 60's \cite{TT,Kumm,Schlot}. However, they lack 
consistency with each other and neglect an important effect of the 
Friedel oscillations. In the present paper we calculate analytically  
corrections to the AG result for the transition temperature resulting 
from local variations of superconducting order parameter induced by 
nonmagnetic and magnetic impurities for various types of pairing. 

Our main results can be summarized as follows. Spatial variations of 
$\Delta$, which appear only at very short distances of the order of 
$k_F^{-1}$ from an impurity site, reduce the pair-breaking effect on 
$T_c$. The correction to the AG result is of the order of $T_c/E_F$ 
and becomes important for short-coherence length superconductors such 
as high-$T_c$ cuprates. In $s$-wave superconductors small amounts of 
nonmagnetic impurities can {\it increase} the transition temperature, 
if other effects, such as an energy dependent density of states, are 
small. The effect of spatial inhomogeneity is enhanced in 
two-dimensional superconductors compared to three-dimensional systems. 

In our treatment of impurities in superconductors we go beyond 
the standard assumption of a spatially constant order parameter
\cite{AGm,AGnm}.
The starting point is the 
linearized gap equation \cite{rev}
\begin{equation}
\Delta({\bf k},{\bf q}) = - T
\sum_{\omega,\bf k',\tilde{k},\tilde{q}} V({\bf k},{\bf k}') 
G^n_{\omega}({\bf k}'+\case{1}{2}{\bf q},
\tilde{\bf k}+\case{1}{2}\tilde{\bf q}) 
G^n_{-\omega}(-{\bf k}'+\case{1}{2}{\bf q},
-\tilde{\bf k}+\case{1}{2}\tilde{\bf q}) 
\Delta(\tilde{\bf k},\tilde{\bf q}) 
\label{lin0}
\end{equation}
for a singlet superconducting order parameter $\Delta({\bf k},{\bf 
q})$ (we suppress spin indices) with $\bf k$ and $\bf q$ related to 
the relative and the center of mass coordinates, respectively. The 
finite-temperature normal state Green's function 
$G^n_{\omega}({\bf k,k}')$ in an external impurity potential depends 
on the two momenta. The retarded electron-electron interaction is 
taken in the form $V({\bf k},{\bf k}')=- g f({\bf k}) f^*({\bf k}')$, 
i.e., we assume pairing according to a nondegenerate one-dimensional 
irreducible representation of the crystal point group and employ a 
cut-off $\omega_c$, $T_c<\omega_c<E_F$, in sums over Matsubara 
frequencies. Consequently, the superconducting gap in the mixed 
momentum-space representation is factorized $\Delta({\bf k},{\bf R})= 
f({\bf k})\Delta({\bf R})$ and Eq.~(\ref{lin0}) is rewritten for a 
function of space coordinate $\Delta({\bf x})$ as 
\begin{equation}
\Delta({\bf x})=g T_c \sum_{\omega}  \int  d{\bf y}  
\left[K_{0\omega}(|{\bf x- y}|)+ K_{1\omega}({\bf x},{\bf y})
\right] \Delta({\bf y})  . 
\label{lin}
\end{equation}
Here, both the gap and the nonuniform part of the kernel $K_{1\omega}$ 
depend on impurity positions ${\bf R}_j$ ($j=1,\ldots,N_i$). The order 
parameter is written in the form 
\begin{equation}
\Delta({\bf x}) = \langle\Delta\rangle \frac{1-F({\bf x})}
{1 - \langle F\rangle} \ ,
\label{multi}
\end{equation}
where $\langle\Delta\rangle = \frac{1}{V}\int d{\bf x} \Delta({\bf 
x})$, $\langle F\rangle = \frac{1}{V}\int d{\bf x}
F({\bf x})$, and the function $F({\bf x})$ 
describes relative variations of the gap in 
space. Integrating the gap equation (\ref{lin}) over the volume and 
assuming the dilute limit ($\langle F\rangle\ll 1$) we have 
\begin{equation}
1 = \frac{gT_c }{V} \sum_{\omega} \int  d{\bf x}d{\bf y} 
 \Bigl\{  
K_{0\omega}(|{\bf x-y}|)+K_{1\omega}({\bf x},{\bf y}) 
  [1- 
 F({\bf y}) + \langle F\rangle] \Bigr\} .
\label{step1}
\end{equation}
The terms without $F$, when averaged over impurity 
positions, give the standard AG result, 
whereas the remaining members describe corrections
due to spatial variations of the gap. In the low impurity 
concentration limit, the net space variation can be considered as a 
sum of the variations due to each impurity: 
$K_{1\omega}({\bf x},{\bf y}) = \sum_{j=1}^{N_i}
\tilde{K}_{1\omega}({\bf x},{\bf y};{\bf R}_j)$, 
$F({\bf y}) = \sum_{j=1}^{N_i}\tilde{F}({\bf y}-{\bf R}_j)$, 
$n_i=N_i/V$. 
Substituting these back into Eq.~(\ref{step1}) we notice that all terms 
$\tilde{K}_{1\omega}({\bf x},{\bf y};{\bf R}_j)
[\tilde{F}({\bf y}-{\bf R}_k)-\langle\tilde{F}({\bf R}_k)\rangle]$ 
with $j\neq k$ 
vanish after averaging over impurity positions. The impurity averaged 
gap equation has the following form
\begin{equation}
\ln\frac{T_{c0}}{T_c} + \psi(\case{1}{2})  -  
\psi(\case{1}{2}+\rho) 
  = \frac{n_i T_c}{N_0}\sum_\omega
\int d{\bf y}\tilde{K}_{1\omega}({\bf x},{\bf y})\tilde{F}({\bf y}) \ , 
\label{step2}
\end{equation}
where the difference of the di-gamma functions $\psi$'s represents the 
AG result: $\rho=1/4\pi T_c\tau$ for nonmagnetic impurities in an 
unconventional superconductor and $\rho=1/2\pi T_c\tau_s$ for magnetic 
impurities in an $s$-wave superconductor, $\tau$ and $\tau_s$ being 
potential and exchange scattering times, respectively;  $N_0$ is the 
Fermi level density of states.  In the dilute limit the correction to the 
AG result is expressed via single impurity characteristics only. 
Similar equations have been earlier obtained in \cite{TT}. Our 
consideration departs, however, now from previous works. 

In order to express spatial variation of the gap function for a single 
impurity we return back to Eq.~(\ref{lin}). The relation (\ref{multi}) 
is now replaced by $\Delta({\bf x}) = \Delta[1-\tilde{F}({\bf x})]$, 
since $\langle\tilde{F}\rangle$ vanishes in the thermodynamic limit 
for one impurity. Neglecting variations of $\Delta$ one the right hand 
side of Eq.~(\ref{lin}) we immediately arrive at 
\begin{equation}
\tilde{F}({\bf x})=-gT_c\sum_\omega\int d{\bf y}
\tilde{K}_{1\omega}({\bf x},{\bf y})\ .
\label{nonselfc}
\end{equation}
This is the so called non-self-consistent approximation for the spatial 
variation of the gap \cite{Schlot,XW} and its validity for short 
distances from the impurity is justified later. Using symmetry of the 
kernel we finally get from Eq.~(\ref{step2}) 
our main result: 
\begin{equation}
T_{c0}-T_c = \delta T_{\text{AG}}(n_i) - \frac{n_iT_{c0}}{gN_0} 
\int d{\bf x}\tilde{F}({\bf x})^2 \ . 
\label{step3}
\end{equation}
In contrast to previous studies \cite{TT}, the above equation 
takes into full account the effect of spatial variations of the gap, 
including the rapid Friedel oscillations of wave-length $k_F^{-1}$
in the vicinity of the impurity. Though Eq.~(\ref{step3}) contains 
only linear in $n_i$ corrections, it holds for both magnetic and 
nonmagnetic impurities and for an arbitrary type of pairing. As is 
seen from Eq.~(\ref{step3}), the qualitative role of the correction is 
always to reduce the pair-breaking effect. In the case of nonmagnetic 
impurities in an $s$-wave superconductor, when $\delta 
T_{\text{AG}}(n_i) \equiv0$, the transition temperature 
increases as a result of spatial variations. 

\paragraph*{Nonmagnetic impurities in $s$-wave superconductors.}
In order to estimate these effects quantitatively we now turn to the 
calculation of the relative gap suppression 
$\tilde{F}({\bf x})$ in particular cases, 
considering first a $\delta$-function impurity potential 
$U({\bf x})=U\delta({\bf x})$ in an 
$s$-wave superconductor. The $2\times 2$ Nambu Green's function in 
superconducting state is expressed via a $t$-matrix as 
$\hat{G}_\omega({\bf x},{\bf x}')=\hat{G}^\Delta_\omega({\bf x},{\bf x}')
 + \hat{G}^\Delta_\omega({\bf x},0) \hat{t}_\omega 
\hat{G}^\Delta_\omega(0,{\bf x}')$,
where $\hat{G}^\Delta_\omega({\bf x},{\bf x}')$ satisfies Gor'kov's 
equation without impurity potential, but with corresponding variations 
of the gap: 
\begin{equation}
\left[i\omega - h_0({\bf x})\tau_3 -\Delta\tau_1\right] 
\hat{G}^\Delta_\omega({\bf x},{\bf x}')  =  \delta({\bf x}-{\bf x}')
  - \tilde{F}({\bf x})\Delta\tau_1
\hat{G}^\Delta_\omega({\bf x},{\bf x}') \ .
\label{Gorkov}
\end{equation}
The function $\tilde{F}({\bf x})$ has to be found from the gap equation
\begin{equation}
\tilde{F}({\bf x})=\frac{gT}{2\Delta}\sum_\omega\text{Tr}\left[
\tau_1\delta\hat{G}_\omega({\bf x},{\bf x})\right] ,   \ \ 
\delta\hat{G}_\omega =  \hat{G}_\omega - \hat{G}^0_\omega  .
\label{gap}
\end{equation}
Here $\hat{G}^0_\omega({\bf k}) = 
[i\omega-\varepsilon_k\tau_3-\Delta\tau_1]^{-1}$, $\varepsilon_k$ 
is the quasiparticle dispersion, and 
$\tau_i$ are Pauli matrices. The expression for the $t$-matrix
is $\hat{t}_\omega = U\tau_3 
[1-\hat{G}^\Delta_\omega(0,0)U\tau_3]^{-1}$, which we approximate by 
$\hat{t}_\omega \approx \alpha\tau_3/(\pi N_0)$ for 
$\alpha=\pi N_0 U\ll 1$. 

The short-distance behavior of $\tilde{F}({\bf x})$ can 
be found approximating
$\hat{G}^\Delta_\omega$ by $\hat{G}^0_\omega$.
This is exactly the non-self-consistent approximation used above.
The real space representation for $\hat{G}^0_\omega$ is 
\begin{equation}
\hat{G}^0_\omega({\bf x}) = 
-\frac{i}{2\Omega} \Bigl\{  (i\omega+\Delta\tau_1)
[f_{\omega+}({\bf x})-f_{\omega-}({\bf x})] 
  \frac{}{} + i\Omega\tau_3
[f_{\omega+}({\bf x})+f_{\omega-}({\bf x})] \Bigr\} \ ,
\label{superG}
\end{equation}
where
$f_{\omega\pm}({\bf x})=-\pi N_0/(k_F x)\,e^{\pm ik_F x}
e^{- \Omega x/v_F}$, $\Omega=\sqrt{\omega^2+\Delta^2}$.
Substituting it back into Eq.~(\ref{gap}) we find
\begin{equation}
\tilde{F}_0({\bf x}) 
= \alpha gN_0 \frac{\sin 2k_F x}{(k_F x)^2}\,\pi T \sum_{\omega}
\frac{e^{-2\Omega x/v_F}}{\sqrt{\omega^2+\Delta^2}} \ .
\label{nonself}
\end{equation}
A similar result was obtained for a finite-radius impurity by Fetter
\cite{Fetter}. Note that the method used in \cite{TT,Schlot}
for a magnetic impurity would give zero instead of (\ref{nonself})
and, consequently, no effect on the transition temperature.
In the limit $T\rightarrow 
T_c$ one can neglect $\Delta$ and find the following analytic 
asymptotes
\begin{equation}
\tilde{F}_0({\bf x}) = \alpha\,\frac{\sin 2k_F x}{(k_F x)^2} \ \ \ \ 
\text{for}\ k_F^{-1}\lesssim x< \tilde{\xi_0}=\xi_0 
\frac{\pi T_c}{\omega_c}  
\label{shorts}
\end{equation}
and $\tilde{F}_0({\bf x})=2\alpha gN_0\sin(2k_F x)/(k_F x)^2 e^{-x/\xi_0}$ 
for $x> \xi_0$, where $\xi_0=v_F/(2\pi T_c)$ is the coherence length. 
The latter long-distance asymptote is incorrect since no 
such rapid oscillations are present in the Ginzburg-Landau regime. The 
correct $1/x$ behavior is recovered in the self-consistent analysis. 
However, for the purpose of calculating the correction (\ref{step3})
we need to know only the short distance expression (\ref{shorts}).
The function $\tilde{F}_0({\bf x})$ falls of to a negligible value within
a few Fermi wave-lengths $k_F^{-1}$. Such rapid relaxation of the gap
has also been seen in numerical simulations \cite{XW,Flatte}.
Accordingly, the integral in Eq.~(\ref{step3}) quickly converges and 
does not depend on the upper limit. Integral's 
value is $4\pi^2\alpha^2/k_F^3$ and, hence, the transition temperature
of an $s$-wave superconductor with nonmagnetic impurities is
\begin{equation}
T_c = T_{c0} + \frac{\pi}{2\tau} \frac{T_{c0}}{E_F} 
\ln\frac{2\omega_c}{\Delta_0} \ ,
\label{Tc-n}
\end{equation}
where $\Delta_0$ is the zero-temperature gap of a pure system and 
$1/2\tau= \alpha^2 n_i/\pi N_0$ is the normal state scattering rate. 
The result (\ref{Tc-n}) does not contradict Anderson's theorem, 
which is valid only for a spatially homogeneous superconducting state 
\cite{Anderson}. 

In order to prove the validity of the non-self-consistent approximation we 
return back to Eq.~(\ref{Gorkov}) and write its solution to the first 
order in $\tilde{F}({\bf x})$:
\begin{equation}
\hat{G}^\Delta_\omega({\bf x},{\bf x}') = \hat{G}^0_\omega({\bf x}- 
{\bf x}') 
 -  \int d{\bf y} 
\hat{G}^0_\omega({\bf x}-{\bf y})\tilde{F}({\bf y}) \Delta \tau_1
\hat{G}^0_\omega({\bf y}-{\bf x}') \ .
\label{1st}
\end{equation}
Substituting this into the gap equation and keeping only terms of 
first order in impurity potential we find after Fourier transformation 
\begin{equation}
\tilde{F}({\bf p})\Bigl\{1+\case{1}{2} gT 
\sum_{\omega,{\bf q}} \text{Tr}\left[\tau_1 
\hat{G}^0_\omega({\bf p}+{\bf q})\tau_1\hat{G}^0_\omega({\bf q})
\right]\Bigr\}=\tilde{F}_0({\bf p}) .
\label{Fp}
\end{equation}
The difference with the non-self-consistent solution $\tilde{F}_0({\bf p})$ 
appears because of the term in the curl brackets with the sum over momenta  
given near $T_c$ by
\begin{equation}
I_\omega = - \sum_{\bf q} \frac{
\omega^2 + \varepsilon_{\bf q}\varepsilon_{\bf p+q}}
{(\omega^2 + \varepsilon_{\bf q}^2)
(\omega^2 + \varepsilon_{\bf p+q}^2)}  
\approx  -  \sum_{\bf q} \frac{1}
{\varepsilon_{\bf q}\varepsilon_{\bf p+q}} .
\label{Iw1}
\end{equation}
The short distance behavior of $\tilde{F}({\bf x})$ corresponds to the 
large-$p$ asymptote for $\tilde{F}({\bf p})$. In this case we neglected 
Matsubara frequencies $\omega$ in the above equation and 
found the remaining integral to be
$I_\omega=m^2/4p$.
The self-consistent correction in Eq.~(\ref{Fp}) is 
$gT\sum_\omega I_\omega =\frac{\pi}{4} gN_0/ (\tilde{\xi}_0 p)$ and is 
small for $\tilde{\xi}_0 p>1$. In the corresponding space region 
$x<\tilde{\xi}_0$ the non-self-consistent solution (\ref{shorts}) is a 
good approximation for the gap variations. Numerical data show 
that this may be an accurate approximation even for a strong impurity 
potential \cite{XW}.

\paragraph*{Magnetic impurities.} We consider a classical spin, which 
interacts with conduction electrons via an exchange Hamiltonian 
$\hat{V} = J\delta({\bf x}) S^i \hat{\sigma}^i$, where 
$\hat{\sigma}^i$ are spin Pauli matrices. In order to express 
appropriately corresponding effects, one should include spin indices 
in Eq.~(\ref{lin0}). Invariance of the singlet superconducting order 
parameter with respect to spin rotations \cite{Balat} suggests, 
however, that only rotationally-invariant part of the impurity 
$t$-matrix produces a non-zero effect. Therefore, we can average 
$\hat{t}_\omega = \hat{V}[1-\hat{G}_\omega(0,0)\hat{V}]^{-1}$ with 
respect to the direction of spin, though it does not imply rapid 
fluctuations of $\bf S$. The averaged $t$-matrix is now diagonal in 
spin indices and is approximated as $\hat{t}_\omega=\alpha^2/(\pi 
N_0)^2 \hat{G}^0_\omega(0,0)$ for $\alpha=JS\pi N_0\ll 1$ 
\cite{Schlot}. Assuming a particle-hole symmetry we have 
$\hat{G}^0_\omega(0,0)= -\pi N_0(i\omega+\Delta\tau_1)/\Omega$. 
Following exactly the same procedure as described above we find for 
short distances $x<\tilde{\xi}_0$ 
\begin{equation}
\tilde{F}_0({\bf x}) = \alpha^2\,\frac{2-\cos 2k_F x}{(k_F x)^2} \ . 
\label{shortm}
\end{equation}
For pair-breaking magnetic impurities there is 
an additional non-oscillatory suppression of the superconducting order 
parameter close to the impurity. 
Note that Eq.~(\ref{shortm}) [as well as Eq.~(\ref{shorts})]
fails at very short distances $x\ll k_F^{-1}$, 
where an exact band structure becomes important (see, e.g., 
\cite{Flatte}). The correction (\ref{step3}) can be nevertheless evaluated 
using the expression (\ref{shortm}) with the lower limit cut-off 
$\sim k_F^{-1}$. The result is 
\begin{equation}
T_c = T_{c0}- \frac{\pi}{4\tau_s} \left( 1 - 5.1\alpha^2 
\frac{T_{c0}}{E_F} \ln\frac{2\omega_c}{\Delta_0} \right), 
\label{Tc-m}
\end{equation}
where $1/2\tau_s=\alpha^2 n_i/\pi N_0$. The qualitative difference
in a way how the gap function relaxes near magnetic and nonmagnetic 
impurities
has no effect on the integral in (\ref{step3}), which is always of
the order of $1/k_F^3$ due to rapid relaxation. The difference 
appears, however, in the dependence of the transition temperature 
correction on the strength of perturbation $\alpha$, showing an 
additional smallness of the effect for magnetic impurities. 

\paragraph*{Nonmagnetic impurities in unconventional superconductors.}
The real space formalism used above becomes inapplicable in this case, 
since local variations of the gap function depend on the exact
nature of the equilibrium gap $\Delta({\bf k})=\Delta f({\bf k})$
and are generally highly anisotropic. Our goal is to show that the 
order of magnitude of the correction to the AG result
remains unchanged compared to the 
case of nonmagnetic impurities in conventional superconductors. For 
that we rewrite all equations in the momentum representation, including 
$\delta\hat{G}_\omega({\bf k,k'})=
\hat{G}^0_\omega({\bf k})\hat{t}_\omega\hat{G}_\omega({\bf k'})$
and the gap equation (\ref{gap}) as $\Delta({\bf q}) = -\case{1}{2}gT
\sum_{\omega,{\bf k}} f^*({\bf k})\text{Tr}[\tau_1
\hat{G}_\omega({\bf k}+\case{1}{2}{\bf q, k}-\case{1}{2}{\bf q})]$.
Near $T_c$, the non-self-consistent expression for the relative 
variation of the gap is
$$
\tilde{F}_0({\bf p})=\frac{\alpha gT}{\pi N_0}\sum_{\omega,{\bf k}}
f^*({\bf k})\frac{ 
\varepsilon_{{\bf k}+\frac{1}{2}{\bf p}}f({\bf k}-
\frac{1}{2}{\bf p})+ 
\varepsilon_{{\bf k}-\frac{1}{2}{\bf p}}f({\bf k}+
\case{1}{2}{\bf p}) } 
{(\omega^2+\varepsilon_{{\bf k}+\frac{1}{2}{\bf p}}^2)   
(\omega^2+\varepsilon_{{\bf k}-\frac{1}{2}{\bf p}}^2)}.   
$$
We further neglect $\bf p$ in the arguments of 
$f({\bf k}\pm\case{1}{2}{\bf p})$ and average $|f({\bf k})|^2$ 
over the Fermi surface; this procedure changes the 
numerical factor but not the order of magnitude. After that 
the integrand becomes the same as for the Fourier transform of 
(\ref{shorts}). Therefore, the correction in Eq.~(\ref{step3}) will be 
of the same order of magnitude and finally we have
\begin{equation}
T_c=T_{c0}-\frac{\pi}{8\tau}\left(1-f\,\frac{T_{c0}}{E_F}\right)
\label{Tc-d}
\end{equation}
with $f\!\!\!\sim\,$3--5; in the BCS theory $T_{c0}/E_F 
\approx 0.12\,a/\xi_0$. We now compare this result with 
the numerical data of Franz {\it et al\/}.\ \cite{Franz}. For $\xi_0/a = 
4.7$ these authors found a 14\% correction to the AG result, which 
corresponds to $f=5.5$. For $\xi_0/a = 2.5$ their result is a 50\% 
correction to the slope of $T_c(n_i)$ at $n_i=0$ or $f=10$. 
Experiment on Zn-substituted Y-123 superconductor ($\xi_0/a\sim 5$)
found a 40\% discrepancy with the 
AG theory \cite{exp2}, implying $f\gtrsim 10$. Somewhat larger values 
for the factor $f$ found experimentally and numerically are 
related to the two-dimensional (2D) nature of the copper oxide 
materials (see below).

\paragraph*{Two-dimensional superconductors.} We consider the case of 
an $s$-wave gap. The effect of nonmagnetic impurities can be 
calculated as in the 3D case with the only change that the 
Green's functions have to be taken in the 2D form: 
$f_{\omega\pm}({\bf x}) = -\pi i N_0 H_0^{(1)} [\pm k_Fx (1 \pm 
\Omega/E_F)^{1/2}]$. Using the asymptotic form of the Hankel function 
of the first kind at $k_F x\gg 1$ we have 
$f_{\omega\pm}({\bf x})=N_0 (2\pi/k_F x)^{1/2}e^{\pm i(k_F x-3\pi/4)}  
e^{-\Omega x/v_F}$. Substituting into Eq.~(\ref{superG}) we 
obtain for $x<\tilde{\xi}_0$ 
\begin{equation}
\tilde{F}_0({\bf x})=\frac{2}{\pi}\,\alpha\,\frac{\cos 2k_F x}{k_F x} \ .
\label{short2d}
\end{equation}
The characteristic feature of the reduced 
dimensionality is that the order parameter relaxes more slowly to 
its bulk value. Because of this the integral in Eq.~(\ref{step3}) 
formally diverges for the above $\tilde{F}({\bf x})$ and 
becomes dependent on 
the upper limit. This yields an additional logarithmic factor 
$\ln(E_F/\Delta_0)\!\sim\,$2--4 to the correction in Eq.~(\ref{Tc-n}) 
enhancing effect of spatial variations near impurities on the 
mean-field $T_c$ in 2D. The previous arguments suggest that this 
result must hold for any symmetry of the gap. 

In conclusion, we have calculated a reduction of the pair-breaking 
effect on the superconducting transition temperature due to local 
variations of the order parameter near impurities. Such an effect is 
important in short-coherence-length superconductors and is further 
enhanced for quasi two-dimensional systems. In singlet 
superconductors, spatial variations of the order parameter 
near magnetic impurities are less important for $T_c$ 
than near nonmagnetic. Our theory 
explains numerical data for a model $d$-wave superconductor 
\cite{Franz} and shows the significance of such effects in high-$T_c$ 
cuprates. We also predict an enhancement of the critical temperature 
of an $s$-wave superconductor with nonmagnetic impurities. This 
prediction can be tested numerically extending previous model 
simulations \cite{Franz,XW} and may have an experimental relevance to 
Nd$_{2-x}$Ce$_x$CuO$_4$ and other electron-doped cuprates, which 
are believed to be layered $s$-wave superconductors. 

This work was supported by the National Science and Engineering 
Research Council of Canada. We are grateful to V. I. Rupasov for 
useful discussions.


\begin{references}
\bibitem[*]{Landau} On leave of absence from L. D. Landau Institute
 for Theoretical Physics, Moscow, Russia.

\bibitem{AGm}
A. A. Abrikosov and L. P. Gor'kov, Sov. Phys. JETP {\bf 12}, 
1243 (1961).

\bibitem{AGnm}
A. A. Abrikosov and L. P. Gor'kov, Sov. Phys. JETP {\bf 8}, 
1090 (1959).

\bibitem{Anderson}
P. W. Anderson, J. Phys. Chem. Solids {\bf 11}, 26 (1959).

\bibitem{dwave}
D. A. Wollman {\it et al\/}., Phys. Rev. Lett. {\bf 71}, 2134 (1993); 
C. C. Tsuei {\it et al\/}., {\it ibid\/}.\  {\bf 73}, 593 (1994). 

\bibitem{AGext}
P. J. Hirschfeld and N. Goldenfeld, Phys. Rev. B {\bf 48}, 4219 (1993); 
P. A. Lee, Phys. Rev. Lett. {\bf 71}, 225 (1993); 
G. Preosti, H. Kim, and P. Muzikar, Phys. Rev. B 
{\bf 50}, 1259 (1994).

\bibitem{exp1}
C. Bernhard {\it et al\/}., Phys. Rev. Lett. {\bf 77}, 2304 (1996). 

\bibitem{exp2}
B. Nachumi {\it et al\/}., Phys. Rev. Lett. {\bf 77}, 5421 (1996). 

\bibitem{exp3}
S. H. Moffat {\it et al\/}., Phys. Rev. B {\bf 55}, 14\,741 (1997). 

\bibitem{Radtke}
R. J. Radtke {\it et al\/}., Phys. Rev. B {\bf 48}, 653 (1993). 

\bibitem{Fehr}
R. Fehrenbacher, Phys. Rev. Lett. {\bf 77}, 1849 (1996). 

\bibitem{Franz}
M. Franz {\it et al\/}., Phys. Rev. B {\bf 56}, 7882 (1997).

\bibitem{TT}
T. Tsuzuki and T. Tsuneto, Prog. Theor. Phys. {\bf 37}, 1 (1967);
J. Heinrichs, Phys. Rev. {\bf 168}, 451 (1968).

\bibitem{Kumm}
R. K\"ummel, Phys. Rev. B {\bf 6}, 2617 (1972).

\bibitem{Schlot}
P. Schlottmann, Phys. Rev. B {\bf 13}, 1 (1976).

\bibitem{rev}
see, e.g., M. Sigrist and K. Ueda, Rev. Mod. Phys. {\bf 63}, 239 (1991).

\bibitem{Fetter}
A. L. Fetter, Phys. Rev. {\bf 140}, A1921 (1965).

\bibitem{XW}
T. Xiang and J. M. Wheatley, Phys. Rev. B {\bf 51}, 11\,721 (1995).

\bibitem{Flatte}
M. E. Flatt\'e and J. M. Byers, Phys. Rev. Lett. {\bf 78}, 3761 (1997);
Phys. Rev. B {\bf 56}, 11\,213 (1997).

\bibitem{Balat}
we do not consider the transition into a spin-$\case{1}{2}$
superconducting state, which appears for a strong exchange coupling, 
see M. I. Salkola, A. V. Balatsky, and J. R. Schrieffer, 
Phys. Rev. B {\bf 55}, 12\,648 (1997).
\end{references}
\end{document}